\documentstyle[aps,preprint]{revtex} 

\author{ Italo Guarneri}
\address{
 Universit\'a di Milano, sede di Como, via Lucini 3, I-22100 Como,\\
Istituto Nazionale di Fisica Nucleare, Sezione di Pavia, via Bassi 6,
 I-27100 Pavia,\\
 Istituto Nazionale di Fisica della Materia, Unit\`a di Milano,
via Celoria 16, I-20134 Milano,\\
}
\title{ 
 On the Dynamical Meaning of Spectral Dimensions}
\date{\today}

\begin{document}

\maketitle

\begin{abstract}

Dynamical Localization theory has drawn attention to general spectral 
conditions which make quantum wave packet diffusion possible, and it 
was found that dimensional properties of the Local Density of States 
play a crucial role in that connection. 
In this paper an abstract result in this vein is 
presented. Time averaging over the trajectory of a wavepacket up to time T defines 
a statistical operator (density matrix). The corresponding entropy increases 
with time  proportional to log T, and the coefficient of proportionality is 
the Hausdorff dimension of the Local Density of States, at least if the latter has 
good scaling properties. In more general cases, we give spectral upper and 
lower bounds for the increase of entropy.

\end{abstract}

\draft 

\smallskip\

\section{Introduction.}

B.V.Chirikov is one of the initiators of Quantum Chaos: the research area 
centered about the basic question, which of the distinctive marks of 
chaotic classical systems survive in the quantal domain. His attention was 
mainly focused on 
dynamical, directly observable aspects, and on 
deterministic diffusion in particular,  
 which is probably the most concrete manifestation 
of chaos in classical hamiltonian systems. The Kicked Rotor - a quantum 
version of the Standard Map, to which Chirikov's name is 
tightly associated - 
revealed that  
quantization tends to suppress classical diffusion\cite{CCIF}
, and brought into light the 
phenomenon of Dynamical Localization, so called in view of its formal 
similarity to  Anderson localization\cite{fish}. From the mathematical 
viewpoint, this 
discovery brought the quantum dynamics of chaotic maps into the realm of 
mathematical localization theory, thus significantly enlarging the scope of 
the latter ; from the physical viewpoint, it established a bridge to Solid 
State Physics, which led to identify the Localization Length as a 
fundamental characteristic scale of quantum dynamics, in the presence of a 
classical chaotic diffusion. The quasi-classical estimate 
for the localization length found by Chirikov and co-workers\cite{cis} 
is based on a 
heuristic argument - the so-called Siberian argument\cite{89} - which is 
in fact a relation between dynamical and spectral properties, built upon the 
Heisenberg relation. 
Crudely handwaving though it may appear to a mathematical eye, the Siberian 
argument has a depth, the exploration of which has led to nontrivial 
mathematical results. Properly reformulated, it shows that quantum 
one-dimensional unbounded diffusion is  only possible, if the spectrum 
(whether of energy or of quasi-energy) is singular; where by diffusion we 
loosely  
mean any type of sub-ballistic propagation, with the spread of the wavepacket 
over the relevant domain (position 
or momentum) increasing with some power of time less than 1. 

Attention was thus drawn to the dynamical implications of singular spectra
\cite{bel}, and 
of singular continuous spectra in particular; for, although pure point 
spectra 
can also give rise to unbounded growth of expectation values of observables, 
due to non-uniform localization of eigenfunctions, they do not lead 
to any unbounded spread if the latter is measured by "intrinsic" quantities 
such as inverse participation ratios or the like.

 Spectral analysis of the kicked rotor reveals a 
qualitative scenario 
somewhat similar to the one appearing with quasi-periodic Schr\"odinger 
operators, such as the Harper (or almost-Mathieu) operator. Its
spectral type sensitively depends on the arithmetic nature of an 
incommensuration parameter linked to the kicking period and to the Planck 
constant. For  
 "typical" irrational values there is evidence of pure-point spectrum; 
on the 
other hand,    
for rational values absolute continuity of the spectrum is proven. 
It follows that  for a set of 
"not too irrational" values, presumably of zero measure, but nevertheless  
of the 2nd Baire category,  there is still a continuous spectrum\cite{me}, 
which is suspected, but not proven, to be singular. This issue is 
closely connected to localization, for, if the latter could be 
proven for a dense set of irrationals, then purely singular continuity 
of the spectrum on  a 2nd category parameter set would follow from Simon's 
Wonderland Theorem\cite{simon}.

Generally speaking, there are two kinds of problems associated with 
singular continuous spectra. The first one is physical: what is their 
physical relevance in general, and in Quantum Chaos in particular. 
The second is mathematical, and includes the analysis 
of their dynamical implications.
 
Concerning the physical relevance of singular continuous spectra, and their 
connection to classical chaos, the situation is still unclear. 
Such spectra  have been proven to occur in models, 
which are directly relevant to mesoscopic physics: e.g., 
electrons in quasi crystals, 
and cristalline electrons in magnetic fields. Still,  
physicists often tend to repress them as mathematical curiosities, 
encouraged in that by their instability: they easily collapse into 
pure point spectra under tiny perturbations. Such arguments can be 
reversed, because  point spectra which lie "infinitely close" to 
singular continuous ones may  be expected to have highly 
nontrivial dynamical properties ; it looks likely that, prior to entering 
the 
final 
localized state, the wavepacket dynamics will display features that may be 
better understood by assimilating the spectrum to a "fractal", much in the 
same way that fractal analysis of certain sets which are {\it not}
 fractal is still 
instructive on not too small geometric scales.

How does the diffusion which is produced by a singular continuous spectrum 
compare with classical chaotic diffusion - when the latter is present in the 
classical limit? This question cannot be posed for the best known 
examples of singular spectra such as the Harper model, because those  
have integrable classical limits. For this reason the Kicked Harper model 
was invented\cite{kh}, which is intermediate between the Harper and the 
Kicked Rotor model , sharing quasi-periodicity with the former and 
classical chaoticity with the latter. It turns out that the time scale 
over which quantum diffusion mimics classical chaotic diffusion is distinct  
from  the one where quantum "fractal" diffusion becomes manifest, 
with qualitative and quantitative differencies from the former\cite{bg}. 
On the grounds of such findings it appears that the two types of diffusion 
bear scarcely any relation to each other. This may reflect a qualitative 
difference between their underlying  dynamical mechanisms: whereas   
quasi-classical diffusion proceeds by excitation of  
all states 
around the 
initial one, "fractal" diffusion is brought about by   
 a coherent chain of quasi-resonant transitions (however, 
the reader should be 
warned that there are certain risks of over-simplification in this 
qualitative picture). In any case, the role of classical chaos in the 
parametric 
"band dynamics" 
which eventually leads to multifractal spectra has been recently 
demonstrated on numerical results 
\cite{theoketz}.

From the mathematical standpoint, some exact results have been proven, 
which connect asymptotic  
dynamics  to spectral quantities related to  "multifractality" 
of the spectrum (or, more precisely, of the LDOS, Local Density of States).
These 
results consist in estimates for the power-law decay of correlations 
\cite{theo}, and in lower bounds for the spread of wave packets
\cite{89,com,last}. These results rest on asymptotic estimates for Fourier 
transforms of fractal measures\cite{89,strich}. The problem of finding 
 {\it upper} bounds ( 
or possibly exact estimates) for the asymptotic spread of wavepackets is 
still 
open (non-rigorous approaches have been implemented, though
\cite{wilk,fred,mant}).
 
Upper bounds appear to require more detailed information concerning 
the specific structure of the Hamiltonian, and its 
(generalized) eigenfunctions. 
Improved lower estimates 
exploiting both spectral and eigenfunction-related information 
have been obtained, heuristically\cite{ketz} and rigorously\cite{kl} - 
the latter on a special model, which has the striking peculiarity of 
displaying ballistic propagation, even with LDOS of arbitrarily small 
positive Hausdorff dimension. In spite of such findings, 
 the search for 
purely spectral bounds is not yet doomed to failure, at least within 
 the 
class of discrete Schr\"odinger operators; for in that class  
there is a somewhat rigid connection 
between spectral measures and eigenfunctions, so that information about 
the latter is certainly encoded in the LDOS itself\cite{her}. 
An additional problem 
with this class of operators is the role of dimensional properties of 
the {\it global} Density of States (DOS), for which there are contrasting 
indications. On the one hand, the DOS can be smooth  
even in the presence of localization; on the other, in some 
quasi-periodic cases with fractal spectra there are 
numerical indications\cite{fred} 
 that transport is tightly determined by DOS,  though 
numerical analysis shows significant differences between   
the multifractal structures of DOS and LDOS even in such cases. 
\cite{bel1}. 

In summary, no exact one-to-one relation has been as yet established  
between spectral dimensions and asymptotic properties of the dynamics; with 
the only exception of the "correlation dimension", which is known to rule 
the decay
of correlations\cite{theo}. In this paper an abstract result 
is proven, which identifies the information dimension with the coefficient 
of logarithmic growth of a dynamically defined entropy, thus providing 
that dimension with a direct dynamical meaning. Indications will also be 
given, that upper bounds should be sought in terms of fractal (box-counting), 
rather than Hausdorff, dimensions; an abstract example will in fact be given, 
of a zero-dimensional LDOS with fractal dimension 1, which leads to 
ballistic propagation. 
   
It is also worth mentioning that the dynamical 
role of 
dimensional properties of singular 
continuous spectra may be an interesting issue in a purely classical context, 
too.
 Classical dynamical systems which have a singular 
continuous  spectrum (in the orthocomplement of constants) are known long 
since to make up a quite large subset in the class of measure-
preserving transformations\cite{knill} ; 
they often lie close to the bottom of the ergodic hierarchy 
, as they may display weak mixing as maximal ergodic property. 
Beyond that, not much is known about their dynamical properties, 
and about the role of spectral dimensions in particular. It is in fact 
difficult to find concrete examples, in which "transport" can be 
meaningfully investigated. Certain substitution systems are 
rigorously known to have a singular continuous spectral component\cite{quef}, 
 but, to the best of the present author's knowledge, no explicit 
example of a classical dynamical system with a purely singular continuous 
spectrum (in the orthocomplement of constants) is rigorously known.  
Good candidates are  certain polygonal billiards\cite{gut}; another 
system, which can be assigned to this class on the 
grounds of numerical evidence, is the "driven spin model"\cite{piko}, 
which is a classical dynamical system that also admits of a quantum 
interpretation. This model is a member in a class of skew-products for which 
singular continuity of the spectrum is established as a generic property
\cite{quef}. Both for the case of billiards, and of driven spins, 
a multifractal analysis of spectral measures has been numerically 
implemented, and results have been dynamically interpreted 
\cite{artu,piko}.   

 In closing this Introduction, it may be necessary to 
underline that it is far  from comprehensive on some  of the  
general issues touched in it, 
 which go beyond the study of 
dynamical implications of dimensional properties of spectra. 

 This study has found 
motivations 
, among others, from the search for  
a quantum counterpart of deterministic diffusion; 
in contributing this 
paper it is a pleasure to acknowledge Boris Chirikov's tutorial explanations 
of the Siberian argument, to which the results presented below can be 
ultimately traced back. 

\section{Entropy of time-averaging, and its growth.}

Consider discrete-time evolution of a 
quantum system with states in a separable Hilbert space ${\cal H}$ 
 : the state at time $t\in Z$ is $\psi
(t)=U^t\psi (0)$, where $U$ is a fixed unitary operator. While the 
discrete-time formulation used here does not set serious restrictions 
on the elaborations below, most of which 
also apply  to a continuous time dynamics, it has the advantage of 
including quantum maps (which can usually
be pictured as one-cycle propagators of periodically driven systems).

Time averages of an observable $A$ following the evolution  of a given 
initial
state $\psi \equiv \psi (0)$ are defined by%
$$
\left\langle A\right\rangle _T=\frac 1T\sum\limits_{s=0}^{T-1}\left\langle
\psi (s)\left| A\right| \psi (s)\right\rangle 
$$
and are in fact statistical averages, $\left\langle A\right\rangle
_T=Tr(\rho (T)A)$, with the density matrix

\begin{equation}
\label{dens}\rho (T)=\frac 1T\sum\limits_{s=0}^{T-1}\left| \psi
(s)\left\rangle {}\right\langle \psi (s)\right| 
\end{equation}

These density matrices are finite rank, positive operators, and to everyone 
of
them is associated the entropy 
\begin{equation}
\label{entropy}
{\cal S}(\rho (T))=Tr(\rho (T)\ln \rho
(T)^{-1})
\end{equation}
which is a measure for the size of the statistical ensemble 
defined by the 
states of the system between times $0$ and $T-1$; 
in the following it will be denoted ${\cal S}(\psi ,T)$. Our aim here is to
estimate the asymptotic growth of ${\cal S}(\psi ,T)$ with $T$ .

The basic tool in getting upper estimates will be the following well-known
result. For $x\in \left[ 0,1\right] $ define $\Theta (x)=-x\ln x$ if 
$x\neq 0
$, $\Theta (0)=0$; then an elementary convexity argument shows that:

{\bf Lemma} . {\sl For any statistical operator }$\rho ,${\sl \ and for any
\ Hilbert base }$B=\left\{ \varphi _n\right\} _{n\in Z},$%
\begin{equation}
\label{lem}
{\cal S}(\rho )\leq \sum\limits_{n\in Z}\Theta (\left\langle
\varphi _n\left| \rho \right| \varphi _n\right\rangle )
\end{equation}

With the statistical operator defined in eqn.(\ref{dens}), $\left\langle
\varphi _n\left| \rho \right| \varphi _n\right\rangle \equiv p_n(T)$ is just
the averaged probability of finding the system in state $\varphi _n$ 
between times $0$ and $T-1$. 
The rhs of (\ref{lem}) is the Shannon entropy of the probability
distribution $p_n(T)$, which in the following will be denoted $S(\psi ,B,T)$.
Since the rank of (\ref{dens}) is at most $T$, the base $B$ can be chosen so
that the sum over $B$ in (\ref{lem}) contains at most $T$ nonzero terms; 
then well-known properties of the Shannon entropy yield the bound ${\cal S}%
(\psi ,T)$ $\leq $ $\ln T$, which is exact, e.g., if  
$U$ has a Lebesgue spectrum in $[0,2\pi]$, because in that case a base $B$
can be found, so that $U$ acts as a shift over $B$. 
 On the opposite extreme, 
if $U$ has a pure point
spectrum, then using in (\ref{lem}) an eigenbase of $U$ we immediately find
that ${\cal S}(\psi ,T)$ remains bounded at all times. Thus we see that
entropy cannot increase with time faster than $\ln T$, and that its actual
increase is related to the degree of continuity of the spectrum.

This qualitative indication will now be turned into an exact result, which 
calls appropriate dimensions into play, as a measure of the "degree 
of continuity". We first review their definitions.

The spectral measure of $\psi $ (also called Local Density of States at $
\psi )$ is the unique measure $d\mu $ on $\left[ 0,2\pi\right] $ such that, 
at all times $t$,

$$
\left\langle \psi \left| U^t\right| \psi \right\rangle
=\int\limits_0^{2\pi}e^{ it\lambda }d\mu (\lambda ) 
$$
The dependence of this measure on the state $\psi $ will be left understood
in the sequel. Various dimensions of the Hausdorff or multifractal type can
be assigned to the measure $d\mu$; the ones we shall use are the upper and
lower Hausdorff dimensions $\dim {}_H^{\pm }(\mu )$, and the fractal
dimension $\dim {}_F(\mu )$, which are defined as follows.

$\dim {}_H^{-}(\mu )$ is the supremum of the set of values $\alpha \in
\left[ 0,1\right] $ such that $\mu (A)=0$ for all Borel sets $A\subseteq
\left[ 0,2\pi\right] $ which have Hausdorff dimension smaller than $\alpha 
.$

$\dim {}_H^{+}(\mu )$ is the infimum of the set of values 
$\alpha \in \left[
0,1\right] $ such that there is a set $A\subseteq \left[ 0,2\pi\right] $ of
Hausdorff dimension $\alpha $, with $\mu (A)=\mu (\left[ 0,2\pi\right] ).$

If the upper and lower dimensions coincide, then the measure is said to have
 exact Hausdorff dimension, given by their common value. Note that
measures which have both a point and a continuous part do not fall in this
class, if the dimension of the continuous part of the measure is positive.

Finally, the fractal dimension is defined as

\begin{equation}
\label{fracd}
\dim {}_F(\mu )=\sup _{0<\epsilon <1}\inf _K\left\{ \dim {}_F(K):\ K\
compact\ s.t.\ \mu (K)>1-\epsilon \right\}  
\end{equation}
where $\dim_F(K)$ is the fractal (box-counting) dimension of $K.$ In
general, $\dim {}_H^{-}(\mu )\leq \dim {}_H^{+}(\mu )\leq \dim {}_F(\mu )$,
but the three dimensions may coincide; such is the case, e.g., when the
measure has, $\mu -$almost everywhere in $\left[ 0,2\pi\right] ,$ a
well-defined , constant scaling exponent\cite{jmp}. In that case their
common value is the same as the information dimension $D(\mu ).$ For such
''exactly scaling'' measures the behavior of entropy is particularly simple:

\smallskip\ 

{\bf Theorem 1.} {\sl If the spectral measure is exactly scaling, with 
dimension }$%
D $, {\sl then} ${\cal S}(\psi ,T)\sim D\ln T$ {\sl asymptotically as} $%
T\rightarrow \infty $.

\smallskip\

This is the central result of this paper: it will be obtained via
Propositions 1-5 below.  Ineq. (\ref{lem}) suggests that upper bounds 
to entropy growth can be
obtained by choosing a suitable base $B$ in the cyclic subspace of $\psi$, 
and then analyzing the growth with time of the Shannon entropy 
$S(\psi ,T,B)$
of the distribution $p_n(T)$ over the base $B$.

The latter is a measure of the ''width'' of the distribution, and
in order to estimate its growth we will estimate the growth of $n_\epsilon
=n_\epsilon (T,B)$, defined as the smallest integer such that the total
probability assigned by $p_n(T)$ to states 
$\left| n\right| \leq n_\epsilon $
be larger than $1-\epsilon ^2.$ To this end we shall use  the following
technical tool:

\smallskip\ 

{\bf Proposition 1}.{\sl \ Let} $B$ {\sl be any base in the cyclic subspace
of} $\psi $, $K_{\epsilon}$ {\sl a compact set in} 
$\left[ 0,2\pi \right] $ {\sl such
that} $\mu (K_{\epsilon})>1-\frac{\epsilon ^2}8$, {\sl and} $N$ 
{\sl an integer} $>1$. 
{\sl Given a partition of }$\left[ 0,2\pi \right] $ {\sl in intervals} 
$%
I_j=\left[ 2\pi jN^{-1},2\pi (j+1)N^{-1}\right] $, 
($j=0,...,N-1),$ {\sl let}

$$
W_B(n,N)=\sum\limits_{I_j\cap K_{\epsilon}\neq \emptyset }
\left| \left\langle \psi
\vert E_{I_j}\varphi _n\right\rangle \right| ^2 
$$
{\sl where} $E_{I_j}$ {\sl are spectral projectors associated with the 
intervals} $I_j $.
 {\sl Define} $\nu _B(\epsilon ,N)$ {\sl as the smallest integer} $%
\nu $ {\sl such that} $\sum_{\left| n\right| >\nu }W_B(n,N)<\epsilon $. 
{\sl 
Then there are numerical constants }$c_1,c_2$ {\sl so that: }

$$
n_\epsilon (T,B)\leq \nu_B (c_2\epsilon ^3, N ) 
$$
{\sl for all} $N>c_1\epsilon^{-1}T$.

{\it Proof :} Note that $\sum\limits_{n\in Z} W_B(n,N)\leq 1$, 
so $\nu_B(\epsilon,N)$ is 
a meaningful quantity. 
First, we use the Spectral theorem to identify the cyclic 
subspace of $\psi$ with $L^2([0,2\pi],d\mu)$, in which  $\psi(t)$ is 
represented by the function $e^{i\lambda t}$ of $\lambda\in[0,2\pi]$. 
Then we define stepwise approximations to $\psi(t)$ for $0\leq t<T$ :

$$
\psi _{K_{\epsilon},N}(t)=\sum\limits_{I_j\cap K_{\epsilon}\neq \emptyset }
e^{2\pi ijtN^{-1}}\chi _{I_j}(\lambda ) 
$$
It is immediately seen that

$$
\left\| \psi (t)-\psi _{K_{\epsilon},N}(t)\right\| ^2\leq 
\frac{\epsilon ^2}8+
\frac{4\pi^2t^2 
}{N^2} 
$$
which can be kept $\leq \frac{\epsilon ^2}4$ at all times from $0$ to $T$,
by choosing $N>c_1 T\epsilon ^{-1}$, with $c_{1}$ a
numerical factor. Then ,

$$
\sum\limits_{\left| n\right| >n_0}p_n(T)\leq \frac{\epsilon ^2}2+\frac
2T\sum_{s=0}^{N-1}\sum\limits_{\left| n\right| >n_0}\left| \left\langle
\varphi _n\vert\psi _{K_{\epsilon},N}(s)\right\rangle \right| ^2 
$$

$$
\leq \frac{\epsilon ^2}2+{\frac {1}{2c_2\epsilon}} \sum\limits_{\left| n\right|
>n_0}\sum\limits_{I_j\cap K_{\epsilon}\neq \emptyset }\left| \left\langle 
\varphi
_n\vert\chi _{I_j}\psi )\right\rangle \right| ^2 
$$

$$
=\frac{\epsilon ^2}2+{\frac{1}{2c_2\epsilon}} \sum\limits_{\left| n\right|
>n_0}W_B(n,N) 
$$
which will be less than $\epsilon ^2$ if $n_0\geq \nu (c_2\epsilon ^3,N)$  
(note that in the 1st inequality the upper limit of the sum over $s$ 
has been changed from $T-1$ to $N-1$, which is certainly larger).$\Box $

{\it Remark.}  Viewed as functions of $n$ at 
fixed $\lambda$ in the spectrum of $U$, 
the functions $\varphi_n(\lambda)$ are (generalized) eigenfunctions of $U$.   
Thus proposition (1) establishes a connection between dynamics and structure 
of eigenfunctions.

If the measure is purely continuous, then there is 
a particular base $B_F$ in the cyclic subspace of $\psi $,
which allows for optimal control on the growth of entropy . 
The  vectors of $B_F$ are  represented by functions 
 $\varphi _n\in $ $L^2(\left[ 0,2\pi\right] ,d\mu )$ defined as follows:

\begin{equation}
\label{lab}
\varphi _n(\lambda )=e^{2\pi inF(\lambda )} 
\end{equation}
where $F(\lambda )=\mu (\left[ 0,\lambda \right] )$ is the distribution
function of the spectral measure. The $\varphi _n$ ($n\in Z$) are a complete
orthonormal set because they are the image of the Fourier base $\left\{
e^{2\pi inx}\right\} _{n\in Z}$ in $L^2(\left[ 0,1\right] ,dx)$ under the
isomorphism which is established (when $d\mu$ is continuous)  
between $L^2(\left[ 0,2\pi \right] ,d\mu )$ and $%
L^2(\left[ 0,1\right] ,dx)$  by $\lambda \rightarrow F(\lambda )$. 

\smallskip\ 

{\bf Proposition 2.} {\sl If } $d\mu$ {\sl is purely continuous then
there is a numerical constant} $c_5$ {\sl 
so that, for any }$d>\dim_F(\mu)$, 
$n_{\epsilon}(T,B_F)\leq c_5\epsilon^{-4}T^d$ {\sl 
for all sufficiently large} $T$.

\smallskip\

{\it Proof.} We use Proposition 1. Observing that 
\begin{equation}
\label{sin}
\left| \left\langle E_{I_j}\varphi _n\vert\psi \right\rangle \right| ^2
=\left|
\int_{I_j}e^{2\pi inF(\lambda )}dF(\lambda )\right| ^2=\frac{\sin {}^2\pi
n\mu (I_j)}{\pi ^2n^2} 
\end{equation}
we obtain
$$
\sum\limits_{\left| n\right| >n_0}W_{B_F}(n,N)\leq c_3\frac{\sharp 
(K_{\epsilon},N)}{n_0} 
$$
where $\sharp (K_{\epsilon},N)$ is the number of intervals $I_j$ which 
overlap $K_{\epsilon}.$ We
now choose $K_{\epsilon}$ so that its box-counting dimension be smaller 
than $\dim {}_F(\mu )
$, which is made possible by the very definition (\ref{fracd})
 of the latter quantity. 
 If we use dyadic partitions,  $N=2^M$, then, for any  
 $d>\dim_F(\mu)$, 
$$
 \sharp(K_{\epsilon},N)<2^{Md}  
$$
for all sufficiently large $M$, so $\nu(\epsilon,2^M)<c_4\epsilon^{-1}
2^{Md}$. Finally, defining $M$  by 
$2^{M-1}\leq c_1 T\epsilon^{-1}<2^{M}$, proposition 1 says that 
$n_{\epsilon}(T)<c_5\epsilon^{-4}T^{d}$ for all sufficiently 
large $T$.
$\Box $

\smallskip\

{\bf Proposition 3.}{\sl If} $d\mu$ {\sl is purely continuous, then} 
$\lim\sup_{T\to\infty}{\frac{S(\psi,B_F,T)}{\ln T}}\leq\dim_F(\mu)$.

\smallskip\

{\it Proof.} Since $S(\psi,B_F,T)$ does not depend on the labeling of 
the base 
vectors, at given time $T$ let us rearrange them in such a way that 
the probability supported by a vector  is monotonically non- increasing 
with 
the label of the vector, thus obtaining a base 
${\overline B}_F$; then clearly $n_{\epsilon}(T,{\overline B}_F)\
\leq n_{\epsilon}(T, B_F)$, so Proposition 2 also holds for 
${\overline B}_F$. 
Therefore, the distribution ${\overline p}_n (T)$ over ${\overline B}_F$ 
obeys
$$
\sum\limits_{m>n}{\overline p}_m(T)\leq c_6T^{\frac {d}{2}}
n^{-{\frac{1}{2}}}
$$
Monotonicity of ${\overline p}_n(T)$ then implies
\begin{equation}
\label{maj}
{\overline p}_n(T)\leq c_7T^{\frac {d}{2}}n^{-{\frac{3}{2}}}
\end{equation}
Let ${\overline n}_{\epsilon}$ the smallest integer larger than 
$c_5\epsilon^{-4}T^d$, so that the total probability on states of 
${\overline B}_F$ beyond ${\overline n}_{\epsilon}$ 
is less than $\epsilon^2$ by construction. For 
$n>{\overline n}_{\epsilon}$ 
the rhs of (\ref{maj}) is certainly smaller than $e^{-1}$ 
for small enough $\epsilon$, so we can use monotonicity of $\Theta(x)$ 
for $x\in(0,e^{-1})$ to the effect that:
$$
\sum\limits_{n}\Theta({\overline p}_n(T))=\left(\sum\limits_{n\leq
{\overline
 n}_{\epsilon}}
+ \sum\limits_{n>{\overline n}_{\epsilon}}\right)
\Theta({\overline p}_n(T))
$$
$$
\leq \ln{\overline n}_{\epsilon}-\ln(1-\epsilon^2)+
\sum\limits_{n>{\overline n}_{\epsilon}}
\Theta(c_7T^{\frac{d}{2}}n^{-{\frac{3}{2}}})
$$
$$
\leq (1+c_8\epsilon^2)\ln{\overline n}_{\epsilon}+ 
O(\epsilon^2\ln{\frac{1}{\epsilon}})
$$
where the last term is only dependent on $\epsilon$. 
 Proposition 2  follows immediately from 
 the definition of  
${\overline n}_{\epsilon}$, because 
$S(\psi, B_F,T)=S(\psi,{\overline B}_F,T)$, and $\epsilon$ 
is arbitrary as well as 
$d>\dim_F(\mu)$.$\Box$
  
Let us extend Proposition 3 to measures 
having a point component, $d\mu=d\mu_p+d\mu_c$. Let 
${\cal H}_p$, ${\cal H}_c$ be the pure point and the continuous subspace 
of the evolution operator $U$, and $P=
\mu_p([0,2\pi])$  the squared norm of the projection of $\psi$ on 
${\cal H}_p$. 
 Then we can write   
$\psi={\sqrt P}\psi_p+{\sqrt{ 1-P}}\psi_c$, with 
$\psi_p\in{\cal H}_p,\psi_c
\in{\cal H}_c$ and 
$\vert\vert\psi_p\vert\vert=\vert\vert\psi_c\vert\vert=1$.

\smallskip\

{\bf Proposition 4.} {\sl If } $d\mu_c$ {\sl denotes the continuous 
component of the 
spectral measure, and} $P$ {\sl 
is the squared norm of the pure point component of} 
$\psi$, {\sl then}
\begin{equation}
\label{fin}
\lim\sup_{T\to\infty}{\frac{{\cal S}(\psi,T)}
{\ln T}}\leq (1-P)\dim_F(\mu_c)
\end{equation}

{\it Proof.} 
Let us choose the base $B=B_p\cup B_c$, where $B_p$ is an eigenbase of $
U$ in ${\cal H}_p$ and $B_c$ is the base (\ref{lab}) associated with 
$d\mu_c$ 
in ${\cal H}_c$. The vectors $\psi_c,\psi_p$ evolve, under repeated 
applications of $U$, independently of each other,with spectral measures 
$d\mu_p,d\mu_c$ respectively; their distributions over $B$ are disjoint, 
so one 
immediately finds that  
that:
\begin{equation}
S(\psi,B,t)=PS(\psi_p,B_p,T)+(1-P)S(\psi_c,B_c,T)+P\ln{\frac{1}{P}}+
(1-P)\ln{\frac{1}{1-P}}
\end{equation}  
The 1st term in the rhs is constant in time, and the second is 
estimated by proposition 2. $\Box$  

The following proposition  sets a  lower bound to entropy. 

\smallskip\

{\bf Proposition 5.} 
$\lim\inf_{T\to\infty}{\frac{{\cal S}(\psi,T)}{\ln T}}
\geq dim^{-}_{H}(\mu)$.

{\it Proof. } This proposition is an immediate consequence, 
not of published results, but of their 
proofs. 

Let $B$ be an eigenbase of $\rho(T)$, so that (\ref{lem}) beomes an 
equality.  Such a base consists of at most $T$ vectors in the subspace 
spanned by $\psi(0),...,\psi(T)$, plus any orthonormal set spanning the 
orthogonal complement of that subspace. A  
 lower bound to $S(\psi,B,T)$ is established as follows. Given $\epsilon\in
(0,1)$ let $m_{\epsilon}(T)$ be the smallest number of base vectors which 
are needed to support more than $1-\epsilon$ of the distribution $p_n(T)$. 
In the Appendix we prove the following elementary estimate, which 
holds if $m_{\epsilon}>3$:

\begin{equation}
\label{lwb}
S(\psi,B,T)\geq \epsilon \ln {\frac{1}{\epsilon}} +
\epsilon \ln (m_{\epsilon}(T)-3)
\end{equation}

General lower bounds on wave packet propagation\cite{jmp} 
entail that, for any small $\eta>0$,
$m_{\epsilon}(T)> c_{\epsilon}T^{\dim^{-}_{H}(\mu)-\eta}$ at all 
sufficiently large 
$T$. Therefore, 
unless $dim^{-}_{H}(\mu)=0$ (in which case Proposition 4 is obviously true), 
 $m_{\epsilon}(T)$ will 
be definitively larger than 3, we can insert the  lower bound on $
m_{\epsilon}(T)$ into (\ref{lwb}), and thus obtain 
the desired result, because $\eta,\epsilon$ are arbitrary. $\Box$
 
\smallskip\

Thus theorem 1 finally emerges, as a consequence of propositions 1-5 for 
the special case that the spectral measure is exactly scaling (proposition 
4  being necessary to that end only in case of  zero-dimensional measures).

\smallskip\

{\it Remarks:} 

\noindent
1. For measures having a point 
component, 
the lower bound $0$ given by proposition 5  is not optimal: one can prove 
that  
$\dim^{-}(\mu)$ can be replaced by $(1-P)\dim^{-}_{H}(\mu_c)$.

\noindent
2. The dynamical entropy defined by (\ref{entropy}) is not related to 
quantum 
analogs of the Kolmogorov-Sinai entropy, as it is only determined by 2-point 
time correlations . In fact,
\begin{equation}
{\cal S}(\psi,T)=\sum\limits_{i}\Theta(p_i)
\end{equation}
where $p_i$ are eigenvalues of $\rho(T)$. It is easily seen that 
$p_i=T^{-1}r_i$, where $r_i$ are eigenvalues of the $T\times T$ 
autocorrelation (Toeplitz) 
matrix $R_{st}=\langle\psi(s)\vert\psi(t)\rangle={\hat\mu}(t-s)$ (the hat   
denoting Fourier transform). In this light, Theorem 1 also applies to 
 $L^2-$ stationary stochastic processes, as a statement relating 
the asymptotic distribution of their autocorrelation eigenvalues to scaling 
properties of the power spectrum. It may result in a convenient method 
for numerically estimating the dimension of the power spectrum.

3. The apparition of  the fractal, rather than the Hausdorff, 
 dimension in the upper bound of Propositions 3,4 
is not an artifact of the proof. In 
the Appendix an abstract example is given of a spectral measure which has 
zero Hausdorff dimension and fractal dimension $1$, for which  
$\lim\sup_{T\to\infty}{\frac{ S(\psi,B_F,T)}{\ln T}}=1$. This "frequently 
ballistic" behavior is also detected in the growth of moments of the 
probability distribution $p_n(T)$, and it is 
controlled by the fractal dimension.

\par\vfill\eject

\section{Appendix.}

\noindent
{1. \it Proof of ineq. (\ref{lwb})}:

Denote  $S(T)=S(\psi,T,B)$ and define ${\cal A}_{\epsilon,T}$ as the set 
of 
labels $n\in Z$ such that $\ln{\frac{1}{p_n(T)}}>S(T)/\epsilon$. From

\begin{equation}
S(T)\geq\sum\limits_{n\in{\cal A}_{\epsilon,T}}\Theta(p_n(T))\geq 
\epsilon^{-1}S(T)\sum\limits_{
n\in{\cal A}_{\epsilon,T}}p_n(T)\end{equation}

it follows that the complement ${\cal B}_{\epsilon,T}$ 
of ${\cal A}_{\epsilon,T}$ 
supports more than $1-\epsilon$ of the distribution $p_n(T)$; hence, 
 it cannot consist of less than 
$ m_{\epsilon}(T)$ elements. 
Now define $\overline{\cal B}_{\epsilon,T}$ as the set of those elements 
of ${\cal B}_{\epsilon,T}$ which have $p_n(T)<e^{-1}$. There cannot be 
less 
than  $m_{\epsilon}-3$ such elements, therefore 

\begin{equation}
S(T)\geq\sum\limits_{n\in\overline{\cal B}_{\epsilon,T}}\Theta(p_n(T))
\geq(m_{\epsilon}-3)\epsilon^{-1}S(T)e^{-S(T)/\epsilon}
\end{equation}
because $\Theta(x)$ is increasing in $(0,e^{-1})$; (\ref{lwb}) immediately 
follows.$\Box$ 

\smallskip\ 

\noindent
{\it 2. An example of a non-exactly scaling, continuous measure, and 
 dynamical consequences thereof.}

The construction below is a special case in a class of measures 
taken from ref.\cite{caltech}. 
Write $\lambda\in[0,2\pi]$ as $2\pi x$ with $x\in[0,1]$, and let 
$a_n(x)\in\{0,1\}$ be the binary digits of $x$. A well-known construction 
allows to define measures on $[0,2\pi]$ as images of cylinder-set measures 
on $\{0,1\}^{\bf N}$, which are in turn constructed by assigning the 
distribution of the random variables $a_n(x)$. Let us consider the 
particular measure $d\mu$ in $[0,2\pi]$ which is obtained when 
the $a_n$'s are independent random variables distributed as follows: 
$a_n(x)=0$ with probability $1$ if $k!\leq n<(k+1)!$ with $k$ even, 
 $a_n(x)=0$ with probability $\frac{1}{2}$ if $k!\leq n<(k+1)!$ with $k$ 
odd.

For integer $k$, consider the dyadic partition of $[0,2\pi]$ in $
2^{k!}$ intervals of equal size $\Delta_k$. It is easily seen that these 
intervals have either measure $0$ or measure $\mu_k$, given by: 
\begin{equation}
\mu_{k}=\prod\limits_{j=1}^{r}\left(\frac{1}{2}\right)^{(2j-1)(2j-1)!}
\quad for\quad even\quad k=2r;\quad
\mu_{k}=\frac{1}{2}\mu_{k-1}\quad for\quad odd\quad k
\end{equation}
whence it follows that
\begin{equation}
\lim_{r\to\infty}{\frac{\ln \mu_{2r+1}}{\ln\Delta_{2r+1}}}=0;
\quad
\lim_{r\to\infty}{\frac{\ln \mu_{2r}}{\ln\Delta_{2r}}}=1.
\end{equation}
Then  $\dim_H(\mu)=0$. In fact, for $\lambda$ in the 
support of $d\mu$, let 
$I_{\delta}(\lambda)=(\lambda-\delta,\lambda+\delta)$.  
If $\delta_k\equiv2\Delta_k$, then,  for 
all $k$, $I_{\delta_k}(\lambda)$ contains the full dyadic interval 
of size 
$\Delta_k$, and measure $\mu_k$, which contains $\lambda$, so that:  
$$
\alpha(\lambda)\equiv\lim\inf_{\delta\to 0}
{\frac {\ln\mu(I_{\delta}(\lambda))}{\ln 2\delta}}
\leq
\lim\inf_{r\to\infty}
{\frac {\ln\mu(I_{\delta_{2r+1}}(\lambda))}{\ln 2\delta_{2r+1}}}
\leq\lim_{r\to\infty}{\frac {\ln \mu_{2r+1}}{\ln 4\Delta_{2r+1}}}=0
$$
(note that ${\ln 4\Delta_{2r+1}}$ is negative at large $r$). Therefore  
 the Hausdorff dimension of $d\mu$ is zero, because it coincides 
with the essential supremum with respect to $d\mu$ of the scaling exponent 
$\alpha(\lambda)$\cite{jmp}.

On the other hand, the fractal dimension (\ref{fracd}) of $d\mu$ is 
$1$.  
In fact, if a compact $K$ has $\mu(K)>1-\epsilon$, then a covering of 
$K$ 
with dyadic intervals of the $(2r)!$-th generation requires at least 
$\sharp_K\geq(1-\epsilon)\mu_{2r}^{-1}$ intervals, so
$$
\dim_F(K)\geq\lim_{r\to\infty}{\frac{\ln\mu_{2r}}{\ln\Delta_{2r}}}=1.
$$
Let us explore how does a wavepacket with the just defined 
spectral measure $d\mu$ 
spread over the base $B_F$ defined by (\ref{lwb}); specifically, we shall 
use (\ref{lab}) to estimate the growth of $S(\psi,B_F,T)$ from below. 

Given $\epsilon\in(0,1)$, let us choose $\epsilon_1<1$ such that 
$1-\epsilon^2-{\frac{\epsilon_1^2}{2}}>0.$ Then, 
going back to the proof of Proposition 1, and using the same notations, 
we have that, for any finite set ${\cal F}$ of indices,
\begin{equation}
\label{app1}
\sum\limits_{n\in{\cal F}}p_n(T)\leq
{\frac{\epsilon_1^2}{2}}+{\frac{1}{2c_2\epsilon_1}}
\sum\limits_{n\in{\cal F}}W_{B_F}(n,N)
\end{equation}
if $N>c_1 T\epsilon_1^{-1}$. For all integer $r$, 
define $T_r^{(\epsilon_1)}$ as the largest 
integer less or equal to $\epsilon_1 c_1^{-1}2^{(2r)!}$, so that 
(\ref 
{app1}) holds  
with $T=T_r^{(\epsilon_1)}$ and $N=N_r=2^{(2r)!}$ for all values of $r$. 
Since all the intervals in 
the partition have either measure $0$ or measure $\mu_{2r}$, 
from (\ref{sin}) we get  
$W_{B_F}(n,N_r)\leq \mu_{2r}$. 
Substituting this into (\ref{app1}) we find that, 
in order that the total probability at time $T_r^{(\epsilon_1)}$ 
on states $n\in{\cal F}$ be larger than  $1-\epsilon^2$, 
${\cal F}$ has to be chosen 
such that
$$
\sharp({\cal F})>2c_2\left(1-\epsilon^2-{\frac{\epsilon_1^2}{2}}\right)
\epsilon_1 
\mu_{2r}^{-1}
$$
Therefore, since  $m_{\epsilon}(T)$ in (\ref{lwb}) 
is the smallest number of base vectors 
which support more than $1-\epsilon^2$ of the distribution $p_n(T)$, 
$$
\lim\sup_{T\to\infty}{\frac{\ln m_{\epsilon}(T)}{\ln T}}
\geq
\lim\sup_{r\to\infty}{\frac{\ln m_{\epsilon}(T_r^{(\epsilon_1)})}
{\ln T_r
^{(\epsilon_1)}}}
\geq
\lim\sup_{r\to\infty}{\frac{\ln \mu_{2r}^{-1}}{\ln T_r^{(\epsilon_1)}}}=
\lim_{r\to\infty}{\frac{\ln \mu_{2r}}{\ln\Delta_{2r}}}=1
$$
 Estimate (\ref{lwb}) and Proposition 2 now yield  
$\lim\sup_{T\to\infty}{\frac{ S(\psi,B_F,T)}{\ln T}}=1$. Much in the same 
way one finds that the growth of $M_q(T)$ (the $q$-th moment of the 
probability 
 distribution $p_n(T)$) follows $\lim\sup_{T\to\infty}{\frac{\ln M_q(T)}
{\ln T}}\geq q$ (note however that in the present case $M_q(t)<\infty$ 
only if 
$q<1$).

\end{document}